\documentclass[11pt]{article}

\usepackage{hyperref,amsfonts,amsmath,amssymb,graphicx,bm,subfigure,a4wide,cite}

\numberwithin{equation}{section}

\begin{document}

\title{\textbf{Magnetic fields in inhomogeneous axion stars}}

\author{Petr Akhmetiev and Maxim Dvornikov\thanks{maxdvo@izmiran.ru}
\\
\small{\ Pushkov Institute of Terrestrial Magnetism, Ionosphere} \\
\small{and Radiowave Propagation (IZMIRAN),} \\
\small{108840 Moscow, Troitsk, Russia}}

\date{}

\maketitle

\begin{abstract}
We study the time evolution of magnetic fields in various configurations of spatially inhomogeneous pseudoscalar fields, which are the coherent superposition of axions. The new induction equation for the magnetic field, which accounts for this inhomogeneity, is derived for such systems. Based on this equation, we study, first, the evolution of two Chern-Simons (CS) waves interacting with a linearly decreasing pseudoscalar field. The nonzero gradient of the pseudoscalar field results in the mixing between these CS waves. Then, we consider the  problem in a compact domain, when an initial CS wave is mirror symmetric. In this situation, the inhomogeneity of a  pseudoscalar field acts as the effective modification of the $\alpha$-dynamo parameter. Thus, we conclude that the influence of a spatially inhomogeneous pseudoscalar field on the magnetic field evolution strongly depends on the geometry of the system.
\end{abstract}



\def\curl{{\operatorname{curl}}}
\def\R{{\Bbb R}}
\def\dd{{\bf{d}}}
\def\b{{\bf b}}
\def\A{{\bf A}}
\def\AA{{\Bbb A}}
\def\B{{\bf B}}
\def\BB{{\Bbb B}}

\section{Introduction}

The most prominent solution of the CP problem in quantum chromodynamics
(QCD) requires the existence of a pseudoscalar particle called the
axion~\cite{PecQui77}. Nowadays axions and axion like particles
are some of the most reliable candidates for dark matter~\cite{DufBib09}.
Despite numerous attempts to directly detect an axion in an experiment,
these particles still remain elusive. The main experimental techniques
used for the axion detection are reviewed in Ref.~\cite{SemYou22}.
The role of axions in astrophysics is highlighted in Ref.~\cite{GalRon22}.

As a rule, axions in the early universe are spatially homogeneous.
However, the coordinate dependence of the axion field is not ruled
out~\cite{Mar16}. It can be the case when the Peccei-Quinn (PQ)
phase transition happens after the reheating during the inflation.
In the present work, we consider such a situation when axions form
spatially confined objects. One of the examples of such structures
are the axion stars~\cite{BarBer11}, which are the solutions of
the wave equation for an axion field in curved spacetime accounting
for a self-interaction of axions. Another possibility for these objects
are axion miniclusters~\cite{KolTka93} which are made of virialized
axions. The characteristics of the axion miniclusters were recently
studied in Refs.~\cite{EnaParSch17,VicRed20}. The impact of the
axions inhomogeneity on the properties of cold dark matter was studied
in Ref.~\cite{KhlSakSok99}.

Axions turn out to interact not only with quarks and between themselves
but also with photons. The most general couplings between axions and
photons are provided, e.g., in Ref.~\cite{SokRin22}. The interactions between axions and other particles within the standard model and beyond are studied in Ref.~\cite{ChoiImShi21}. The mutual evolution
of axions and primordial magnetic fields was studied in Refs.~\cite{LonVac15,DvoSem20}.
The instabilities in the axion MHD were discussed in Ref.~\cite{HwaNoh22}.
We studied the interaction between inhomogeneous axions and helical primordial
magnetic fields in Ref.~\cite{Dvo22}. We assumed in Ref.~\cite{Dvo22}
that the spatial inhomogeneity of axions was isotropic, i.e. the mean
value of the wavefunction gradient vanishes whereas the Laplace operator
survives.

The interaction of axions and magnetic fields results in the magnetic field instability. Therefore, a seed magnetic field can be amplified by an axion dynamo. This process inside a neutron star was recently studied in Ref.~\cite{Anz23}. Various types of the electromagnetic radiation, like fast radio bursts, emission of gamma rays etc., in collisions of axion stars with different astrophysical objects are reviewed in Ref.~\cite{BraZha19}.

In the present work, based on the results of Ref.~\cite{Dvo22},
we consider the mutual evolution of a magnetic field and axions having
the fixed spatial distributions. It can correspond, e.g., to an axion
star. Our analysis is based on the new induction equation for the magnetic field, which is derived for the first time in the present work. This equation accounts for the dependence on spatial coordinates of the classical pseudoscalar field, which is a coherent superposition of axions. Our main goal is to analyze how this inhomogeneous field influences the time evolution of large scale magnetic fields.
For this purpose, we consider two situations: the simplified
one dimensional model, when the pseuduscalar field depends on one spatial coordinate, and the more sophisticated geometry based on the Hopf fibration.

Magnetic fields on 3D sphere relate with the Hopf fibration and are considered in numerous works (see, e.g., Ref.~\cite[Ch.~III]{A-Kh} and references therein).
In Ref.~\cite{A-C-S}, a simplest hyperbolic analogue of the Hopf fibration is applied to construct the magnetic equilibrium on the 3D sphere with the variations of the magnetic permeability. Such a hyperbolic Hopf fibrations are based on the Ghys-Dehornoy examples of geodesic flows.

In Ref.~\cite{K}, the Hopf fibration is used to
translate solutions in a compact domain to solutions in the standard 3D space.
We develop this task in Appendix~\ref{sec3}. Our
approach is different from that in Ref.~\cite{K}, where the magnetic vector potential $\A$ is transformed. We use the Kelvin transformation to construct the magnetic field $\BB$ in the Euclidean space with
various magnetic permeabilities and regular boundary conditions at the infinity.
This approach keeps the magnetic helicity
and admits magnetic energy calculations analogously to Ref.~\cite{A-C-S}.
Because of various magnetic permeabilities, the operator $\curl$ in the Euclidean space is a non-standard and calculations are different with respect to Sec.~\ref{sec:1D}.

In our present construction, the configuration is elliptic, i.e. a scalar curvature parameter in a compact domain is positive, and higher invariants of magnetic lines are unnecessary. 
Using the idea from Ref.~\cite{H-M}, the higher invariants of magnetic lines are required for  fields with the $SU(2)$ symmetry.

Because the initial Chern-Simons (CS) configuration in this domain is mirror symmetric, i.e. the linking numbers for an arbitrary pair of magnetic lines equal to zero, the helicity flow is
completely characterized by a flow of the dispersion of the asymptotic ergodic Hopf invariant.

In Ref.~\cite{A-V}, the analytic formula for the dispersion of the ergodic asymptotic  Hopf invariant, or, the magnetic helicity density is proposed. We recall that the Hopf invariant~\cite{A-Kh} is a density of linking numbers of pairs of magnetic lines.  
The magnetic helicity density has the dimension $\text{G}^2\text{cm}^{-2}$ and is distributed over 4-dimensional configurations. Initial points of magnetic lines in a pair are translated by a prescribed magnetic flow independently. Thus, the Hopf invariant is distributed at transverse sections of pairs of magnetic lines. However, the
dispersion of the Hopf invariant is insufficiently studied since its analytic expression involves an infinite-dimensional space of jets. It is not represented by a  finite-dimensional integral~\cite{Ku}.
A concept of the asymptotic ergodic Hopf invariants assumes that magnetic lines could be non-closed and integrals over a magnetic flow are considered for quasi-periodic functions. For an infinite-dimensional space of jets additional mathematics is required. In Ref.~\cite{G-L}, an approach for non-periodic observables is developped.

The dispersion of the magnetic helicity density, as it is shown in Ref.~\cite{A-V}, has the dimension $\text{G}^4\text{cm}^{-4}$. 
It is distributed over the $2$-points configuration space, and is given by a sequence. The properties of this sequence has to be investigated. It is clear, that the dispersion of the asymptotic Hopf invariant can be used to describe a broken mirror-symmetry at the initial stage of evolution of a nonlinear oscillator given by the interaction with a pseudoscalar field. This problem will be considered elsewhere. 

The present work is organized in the following way. In Sec.~\ref{sec:AXELECTR}, we briefly remind how magnetic fields evolve under the influence of an inhomogeneous pseudoscalar field and derive the new induction equation. We consider the simple one dimensional model in Sec.~\ref{sec:1D}. Then, in Secs.~\ref{sec:3D} and~\ref{sec:HARM}, we qualitatively study a more complicated 3D case. Finally, we conclude in Sec.~\ref{sec:CONCL}. The Kelvin transformation is considered in Appendix~\ref{sec3}. 

\section{Electrodynamics in presence of a classical pseudoscalar field\label{sec:AXELECTR}}

The equations for the interacting electromagnetic and pseudoscalar fields in a curved spacetime have
the form,
\begin{align}
  \frac{1}{\sqrt{-g}}\partial_{\nu}(\sqrt{-g}F^{\mu\nu})+g_{a\gamma}\partial_{\nu}\varphi\tilde{F}^{\mu\nu}+J^{\mu} & =0,
  \label{eq:Maxcov1}
  \\
  \frac{1}{\sqrt{-g}}\partial_{\nu}(\sqrt{-g}\tilde{F}^{\mu\nu}) & =0,
  \label{eq:Maxcov2}
  \\
  \frac{1}{\sqrt{-g}}\partial_{\mu}\left(\sqrt{-g}\partial^{\mu}\varphi\right)+m^{2}\varphi+
  \frac{g_{a\gamma}}{4}F_{\mu\nu}\tilde{F}^{\mu\nu} & =0,
  \label{eq:Maxcov3}
\end{align}
where $F_{\mu\nu}$ is the electromagnetic field tensor, $\tilde{F}^{\mu\nu}=\tfrac{1}{2}E^{\mu\nu\alpha\beta}F_{\alpha\beta}$,
$E^{\mu\nu\alpha\beta}=\tfrac{1}{\sqrt{-g}}\varepsilon^{\mu\nu\alpha\beta}$
is the covariant antisymmetric tensor, $\varepsilon^{0123}=+1$, $g=\det(g_{\mu\nu})$,
$g_{\mu\nu}=\text{diag}(1,-a^{2},-a^{2},-a^{2})$ corresponds to the
Friedmann-Robertson-Walker (FRW) metric with the scale factor $a(t)$,
$g_{a\gamma}$ is the coupling constant, $J^{\mu}=(\rho,\mathbf{J}/a)$
is the external current, $\varphi$ is the classical pseudoscalar field, and
$m$ is its mass.

The system in Eqs.~\eqref{eq:Maxcov1}-\eqref{eq:Maxcov3} is analogous to the axion electrodynamics (see, e.g., Ref.~\cite{LonVac15}) if $\varphi$ is considered as a microscopic field. In our work, we shall study macroscopic spatial distributions of $\varphi$. For example, we can treat $\varphi$ as a coherent superposition of multiple axions.

Using the conformal variables~\cite{BraEnqOle96}, $\mathbf{E}_{c}=a^{2}\mathbf{E}$,
$\mathbf{B}_{c}=a^{2}\mathbf{B}$, $\rho_{c}=a^{3}\rho$, and $\mathbf{J}_{c}=a^{3}\mathbf{J}$,
and applying the results of Ref.~\cite{Dvo22}, we derive the equation
for $\mathbf{B}_{c}$,
\begin{equation}\label{eq:Indeqgen}
  \mathbf{B}'_{c}=\nabla\times
  \left[
    \mathbf{b}\times(\nabla\times\mathbf{B}_{c})+\alpha\mathbf{B}_{c}-\eta_{m}(\nabla\times\mathbf{B}_{c})
  \right],
\end{equation}
where $\mathbf{b}=g_{a\gamma}\nabla\varphi/\sigma_{c}^{2}$, $\alpha=g_{a\gamma}\varphi'/\sigma_{c}$
is the analogue of the $\alpha$-dynamo parameter, $\eta_{m}=\sigma_{c}^{-1}$
is the magnetic diffusion coefficient, $\sigma_{c}\approx10^{2}T_{\mathrm{CMB}}$
is the conformal conductivity of ultrarelativistic plasma, $T_{\mathrm{CMB}}=2.7\,\text{K}$
is the current temperature of the cosmic microwave background radiation, and
the prime means the derivative with respect to the conformal time
$\eta$ defined by $\mathrm{d}t=a\mathrm{d}\eta$. For example, we
can choose $a=T_{\mathrm{CMB}}/T$ and $\eta=\tilde{M}_{\mathrm{Pl}}T_{\mathrm{CMB}}^{-1}(T^{-1}-T_{\mathrm{QCD}}^{-1})$,
where $\tilde{M}_{\mathrm{Pl}}=M_{\mathrm{Pl}}/1.66\sqrt{g_{*}}$,
$M_{\mathrm{Pl}}=1.2\times10^{19}\,\text{GeV}$ is the Planck mass,
and $g_{*}=17.25$ is the number of the relativistic degrees of freedom
at the QCD phase transition~\cite{Hus16}, which happens at $T_{\mathrm{QCD}}\approx100\,\text{MeV}$.
In this case $\eta(T_{\mathrm{QCD}})=0$ and $a_{\mathrm{now}}\equiv a(T_{\mathrm{CMB}})=1$.

Note that the modified induction Eq.~\eqref{eq:Indeqgen} has never been discussed before. It should be compared with the usual MHD induction equation,
\begin{equation}\label{eq:Faraday}
  \mathbf{B}'_{c}=\nabla\times
  \left[
    (\mathbf{v}\times\mathbf{B}_{c})+\alpha\mathbf{B}_{c}-\eta_{m}(\nabla\times\mathbf{B}_{c})
  \right],
\end{equation}
where $\mathbf{v}$ is the macroscopic plasma velocity. Note that $\mathbf{B}_{c}$ in Eqs.~(\ref{eq:Indeqgen}) and~\eqref{eq:Faraday} always has
the zero divergence, $(\nabla\cdot\mathbf{B}_{c})=0$. To derive Eq.~(\ref{eq:Indeqgen}), we keep only the terms linear in $\varphi$.

In order to close the system, we add to Eq.~(\ref{eq:Indeqgen})
the equation for the evolution of $\varphi$ which has the form,
\begin{equation}\label{eq:axconf}
  \varphi^{\prime\prime}+2H\varphi'-\nabla^{2}\varphi+a^{2}m^{2}\varphi=\frac{g_{a\gamma}}{a^{2}}(\mathbf{E}_{c}\mathbf{B}_{c}),
\end{equation}
where $H=a'/a$ is the Hubble parameter. We study $\varphi$ after
the QCD phase transition. If $\varphi$ is a QCD axion, its mass is independent of the plasma
temperature in this case.

In Ref.~\cite{Dvo22}, we considered the electrodynamics of inhomogeneous
axions by assuming that the spatial dependence of their wavefunction is isotropic,
i.e. we supposed that only even number of derivatives, like $\partial_{i}\partial_{j}\varphi$
etc., is nonzero. Now, our task is to study the impact of the term $\mathbf{b}\propto\nabla\varphi$
in Eq.~(\ref{eq:Indeqgen}) on the evolution of the magnetic
field. For this purpose, we study a spatially confined structure
similar to an axion star~\cite{Vic21}.

\section{One dimensional model\label{sec:1D}}

We consider the situation when we have a spherically symmetric distribution of $\varphi$. We direct the $z$-axis along the radius and neglect the dependence on angular coordinates. Hence, both $\mathbf{B}_c$ and $\varphi$ depend on one spatial coordinate $z$, i.e. we consider a 1D model. We take that the magnetic field is the superposition
of two CS waves along the $z$-axis,
\begin{equation}\label{eq:Bpm}
  \mathbf{B}_{c}=\mathbf{B}_{+}+\mathbf{B}_{-},
  \quad
  \mathbf{B}_{+}=B_{+}^{(0)}(\sin kz,\cos kz,0),
  \quad
  \mathbf{B}_{-}=B_{-}^{(0)}(\cos kz,-\sin kz,0),
\end{equation}
where the amplitudes are functions of conformal time $B_{\pm}^{(0)}=B_{\pm}^{(0)}(\eta)$
and $k$ is the wave vector characterizing the scale of the system
$\propto k^{-1}$. Note that $(\mathbf{B}_{+}\cdot\mathbf{B}_{-})=0$,
i.e. these waves correspond to different polarizations. As mentioned above, $\nabla\varphi$ is along the $z$-axis, i.e. $\mathbf{b}=(0,0,b)$.
Thus, we get the following system of differential equations:
\begin{equation}\label{eq:Bpmeq}
  B_{+}^{(0)\prime}=k
  \left[
    kbB_{-}^{(0)}+B_{+}^{(0)}(\alpha-\eta_{m}k)
  \right],
  \quad
  B_{-}^{(0)\prime}=k
  \left[
    -kbB_{+}^{(0)}+B_{-}^{(0)}(\alpha-\eta_{m}k)
  \right],
\end{equation}
for the amplitudes of CS waves. We can see in Eq.~(\ref{eq:Bpmeq})
that the nonzero gradient of the pseudoscalar field $b\propto\partial_{z}\varphi$
mixes the independent CS waves.

The distribution of axions inside an axion star can be quite
sophisticated (see, e.g., Ref.~\cite{Vic21}). We adopt the simple model, in which the
pseudoscalar field has the form,
\begin{equation}\label{eq:linearax}
  \varphi(z,\eta)=
  \begin{cases}
    \varphi_{0}(\eta), & 0<z<R,\\
    \varphi_{0}(\eta)\left(1-\frac{z-R}{\Delta}\right), & R<z<R+\Delta,\\
    0, & z>R+\Delta,
  \end{cases}
\end{equation}
where $R$ is the radius of the axion star core, $\Delta$ is the
depth of the stellar crust, which is supposed to be thin, $\Delta\ll R$,
and $\varphi_{0}(\eta)$ is the oscillating amplitude of the wavefunction.
It means that we consider the analogue of the $\alpha$-dynamo in
a thin layer (see, e.g., Ref.~\cite{Cha20}). Using the fact that
$\mathbf{E}_{c}=\mathbf{J}_{c}/\sigma_{c}=(\nabla\times\mathbf{B}_{c})/\sigma_{c}$
and Eq.~(\ref{eq:Bpm}), we get that $(\mathbf{E}_{c}\cdot\mathbf{B}_{c})=k\left(B_{+}^{(0)2}+B_{-}^{(0)2}\right)/\sigma_{c}$
in Eq.~(\ref{eq:axconf}).

Based on Eq.~(\ref{eq:linearax}), we obtain that $b=-g_{a\gamma}\varphi_{0}/\Delta\sigma^{2}$
and $\nabla^{2}\varphi\equiv\partial_{z}^{2}\varphi=0$. Finally,
Eqs.~(\ref{eq:axconf}) and~(\ref{eq:Bpmeq}) are rewritten in the
form,
\begin{align}\label{eq:sysdim}
B_{+}^{(0)\prime}= & \frac{k}{\sigma_{c}}\left[-\frac{g_{a\gamma}k\varphi_{0}}{\Delta\sigma_{c}}B_{-}^{(0)}+B_{+}^{(0)}\left(g_{a\gamma}\varphi'_{0}-k\right)\right],\nonumber \\
B_{-}^{(0)\prime}= & \frac{k}{\sigma_{c}}\left[\frac{g_{a\gamma}k\varphi_{0}}{\Delta\sigma_{c}}B_{+}^{(0)}+B_{-}^{(0)}\left(g_{a\gamma}\varphi'_{0}-k\right)\right],\nonumber \\
\varphi''_{0}= & -2H\varphi'_{0}-a^{2}m^{2}\varphi_{0}+\frac{g_{a\gamma}k}{a^{2}\sigma_{c}}\left(B_{+}^{(0)2}+B_{-}^{(0)2}\right).
\end{align}
To derive Eq.~(\ref{eq:sysdim}) we take that we are at the bottom
of the stellar crust, i.e. $z\gtrsim R$.

Using the dimensionless variables
\begin{equation}
B_{\pm}^{(0)}=\frac{k}{g_{a\gamma}}\mathcal{B}_{\pm},\quad\varphi_{0}=\frac{\sigma_{c}}{kg_{a\gamma}}\Phi,\quad\eta=\frac{\sigma_{c}}{k^{2}}\tau,
\end{equation}
we rewrite Eq.~(\ref{eq:sysdim}) in the form,
\begin{align}
  \frac{\mathrm{d}\mathcal{B}_{+}}{\mathrm{d}\tau}= & -K\Phi\mathcal{B}_{-}+\mathcal{B}_{+}\left(\xi\Psi-1\right),
  \label{eq:Bpeq}
  \\
  \frac{\mathrm{d}\mathcal{B}_{-}}{\mathrm{d}\tau}= & K\Phi\mathcal{B}_{+}+\mathcal{B}_{-}\left(\xi\Psi-1\right),
  \label{eq:Bmeq}
  \\
  \frac{\mathrm{d}\Psi}{\mathrm{d}\tau}= & -\beta\Psi-\mu^{2}a^{2}\Phi+\frac{1}{a^{2}}
  \left(
    \mathcal{B}_{+}^{2}+\mathcal{B}_{-}^{2}
  \right),
  \label{eq:dotPsi}
\end{align}
where $\Psi=\partial_{\tau}\Phi$, $\mu=m\sigma_{c}/k^{2}$ is the
dimensionless mass of $\varphi$, $K=(\Delta k)^{-1}$, and $\beta=2H\sigma_{c}/k^{2}$.
The terms $\mathcal{B}_{\pm}\Psi$ in the right hand side of Eqs.~(\ref{eq:Bpeq})
and~(\ref{eq:Bmeq}) are responsible for the dynamo amplification
of the magnetic field, i.e. the magnetic field becomes unstable. That
is why, following Ref.~\cite{GruDia94}, we introduce the quenching
factor $\xi=\left[1+(\mathcal{B}_{+}^{2}+\mathcal{B}_{-}^{2})/\mathcal{B}_{\mathrm{eq}}^{2}\right]^{-1}$
in these terms. Here, $\mathcal{B}_{\mathrm{eq}}$ is the equipartition
magnetic field.

The numerical solution of Eqs.~(\ref{eq:Bpeq})-(\ref{eq:dotPsi})
requires the initial conditions. First, we establish the initial condition
for the pseudoscalar field. The energy-momentum tensor of $\varphi$ is
\begin{equation}\label{eq:Tmunu}
T_{\mu\nu}=\partial_{\mu}\varphi\partial_{\nu}\varphi-\frac{g_{\mu\nu}}{2}(g^{\lambda\rho}\partial_{\lambda}\varphi\partial_{\rho}\varphi-m^{2}\varphi^{2}).
\end{equation}
Using Eq.~(\ref{eq:Tmunu}), we get that the total energy
density of $\varphi$ is
\begin{equation}\label{eq:axen}
  \rho_{a}=T_{00}=\frac{1}{2}\left[\dot{\varphi}^{2}+\frac{1}{a^{2}}(\nabla\varphi)^{2}+m^{2}\varphi^{2}\right].
\end{equation}
We suppose that $\dot{\varphi}=0$ initially. Thus, the initial energy
density is
\begin{equation}\label{eq:endensgen}
  \rho_{a}^{(0)}\approx\frac{\varphi_{0}^{2}}{2a^{2}\Delta^{2}}(1+m^{2}a^{2}\Delta^{2}).
\end{equation}
We can compare this quantity with $(\varepsilon mf_{a})^{2}$, where
the factor $\varepsilon\sim10^{-10}$ for a dilute star and $\varepsilon\sim1$
for a dense one (see, e.g., Ref.~\cite{BaiHam18}). Thus, we get
the initial condition for the axion field in terms of the dimensionless
variables
\begin{equation}
  \Phi(T=T_{\mathrm{QCD}})=\frac{\alpha_{\mathrm{em}}\varepsilon ma\Delta k}{\pi\sigma_{c}\sqrt{2(1+m^{2}a^{2}\Delta^{2})}},
  \quad
  \Psi(T=T_{\mathrm{QCD}})=0.
\end{equation}
Here, we use the relation, $g_{a\gamma}\approx\tfrac{\alpha_{\mathrm{em}}}{2\pi f_{a}}$,
where $\alpha_{\mathrm{em}}=7.3\times10^{-3}$ is the fine structure
constant and $f_{a}$ is the PQ constant.

We suppose that the stellar crust has the width $\Delta=0.1R$. The
parameter $k$ in Eq.~(\ref{eq:Bpm}) is related to radius of the
core as $k=R^{-1}.$ It means that $K=10$ in Eqs.~(\ref{eq:Bpeq})-(\ref{eq:dotPsi}).
We take that $k=10^{-8}T_{\mathrm{CMB}}$, which is much less that
the reciprocal Debye scale $k_{\mathrm{D}}=10^{-1}T_{\mathrm{CMB}}$.
The physical size of such a star, if it would evolve to the present
time, is $\sim85\,\text{km}$.

The mass of $\varphi$ is taken to be $m=10^{-3}\,\text{eV}$, which is below
the upper bound established in Ref~\cite{Bus22}. Moreover, we use
the approximate relation (see, e.g., Ref.~\cite{Cha21})
\begin{equation}
  \left(
    \frac{m}{10^{-6}\,\text{eV}}
  \right)
  \approx
  5.7
  \left(
    \frac{f_{a}}{10^{12}\,\text{GeV}}
  \right)^{-1},
\end{equation}
between the axion mass and the PQ constant.

We suppose that the conformal seed magnetic field is $B_{+}^{(0)}(T=T_{\mathrm{QCD}})=4.4\times10^{13}\,\text{G}$
and $B_{-}^{(0)}(T=T_{\mathrm{QCD}})=0$, i.e. only one CS wave in
Eq.~(\ref{eq:Bpm}) is present initially. We use the seed strength
equal to the Schwinger value $B_{\mathrm{crit}}=m_{e}^{2}/e$. The
value of the equipartition field is taken in the range $\mathcal{B}_{\mathrm{eq}}\lesssim10^{-5}$.

\begin{figure}
  \centering
  \subfigure[]
  {\label{fig:Baxionsa}
  \includegraphics[scale=.35]{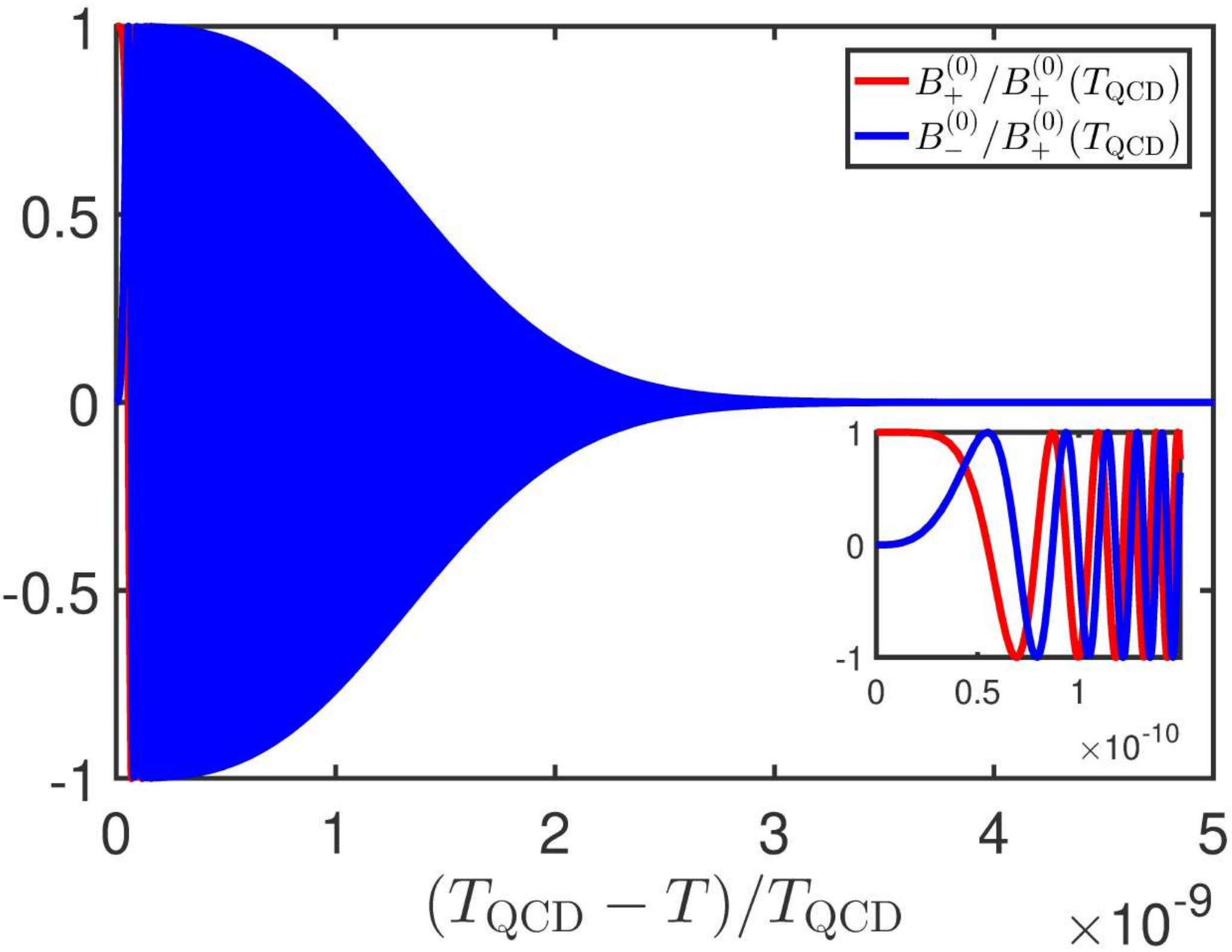}}
  \subfigure[]
  {\label{fig:Baxionsb}
  \includegraphics[scale=.35]{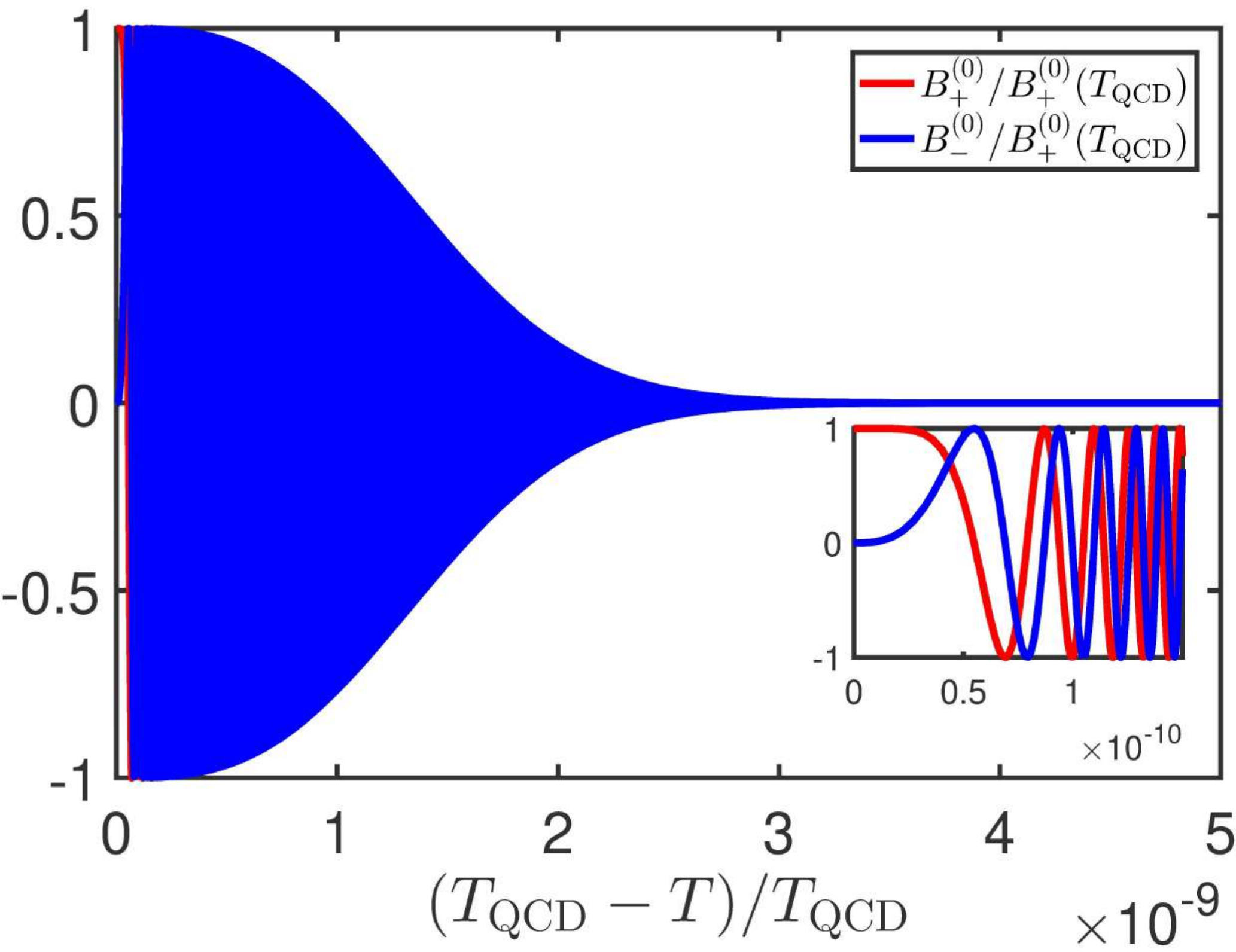}}
  \protect 
\caption{The evolution of $B_{\pm}^{(0)}$ in primordial plasma after the QCD
phase transition. Red and blue lines correspond to $B_{+}^{(0)}$
and $B_{-}^{(0)}$ normalized by $B_{+}^{(0)}(T_{\mathrm{QCD}})$.
The insets show the behavior of $B_{\pm}^{(0)}$ at short evolution
time. The parameters of the system are $m=10^{-3}\,\text{eV}$, $\Delta=0.1R$,
$k=10^{-8}T_{\mathrm{CMB}}$, $B_{+}^{(0)}(T_{\mathrm{QCD}})=4.4\times10^{13}\,\text{G}$
and $B_{-}^{(0)}(T_{\mathrm{QCD}})=0$. Panel (a) corresponds to a
dense star; panel (b) to a dilute one.\label{fig:Baxions}}
\end{figure}

We remind that our main goal is to study the behavior of the magnetic
field under the influence of a pseudoscalar field with a nonzero $\nabla\varphi$.
That is why, first, we plot the amplitudes of CS waves $B_{\pm}^{(0)}$
versus the temperature of primordial plasma $T$ for a dense axion
star in Fig.~\ref{fig:Baxionsa} and for a dilute one in Fig.~\ref{fig:Baxionsb}. In Fig.~\ref{fig:Baxions}, red lines correspond to $B_+^{(0)}$ and blue ones to $B_-^{(0)}$. Both curves are normalized to the initial value $B_+^{(0)}(T_\mathrm{QCD})$ which is nonzero.
We can see that the cases of dense and dilute stars coincide qualitatively.

Magnetic fields in Fig.~\ref{fig:Baxions} oscillate very rapidly even in a short time interval corresponding to $(T_\mathrm{QCD} - T) < 10^{-9} T_\mathrm{QCD}$. That is why curves corresponding to $B^{(0)}_{\pm}$ almost coincide. The behavior of $B^{(0)}_{\pm}$ at $T \lesssim T_\mathrm{QCD}$ is demonstrated in the insets in
Fig.~\ref{fig:Baxions}. One can see in these insets that the numerical solution of Eqs.~\eqref{eq:Bpeq}-\eqref{eq:dotPsi} satisfies the initial condition, $B^{(0)}_{+}(T_\mathrm{QCD}) \neq 0$ (red curve) and $B^{(0)}_{+}(T_\mathrm{QCD}) = 0$ (blue curve). Moreover, we can see a nonzero $\nabla\varphi$ mixes magnetic fields in two different CS waves. This effect is qualitatively predicted earlier by analyzing Eq.~\eqref{eq:Bpmeq}.

We also note that the amplitudes of oscillations of $B^{(0)}_{\pm}$ are almost unchanged in the insets in Fig.~\ref{fig:Baxions}, whereas these functions vanish at greater evolution times. It happens since the magnetic diffusion term is ineffective in short time intervals. At greater evolution times it results in attenuation of magnetic fields.

\begin{figure}
  \centering
  \subfigure[]
  {\label{fig:Benerga}
  \includegraphics[scale=.35]{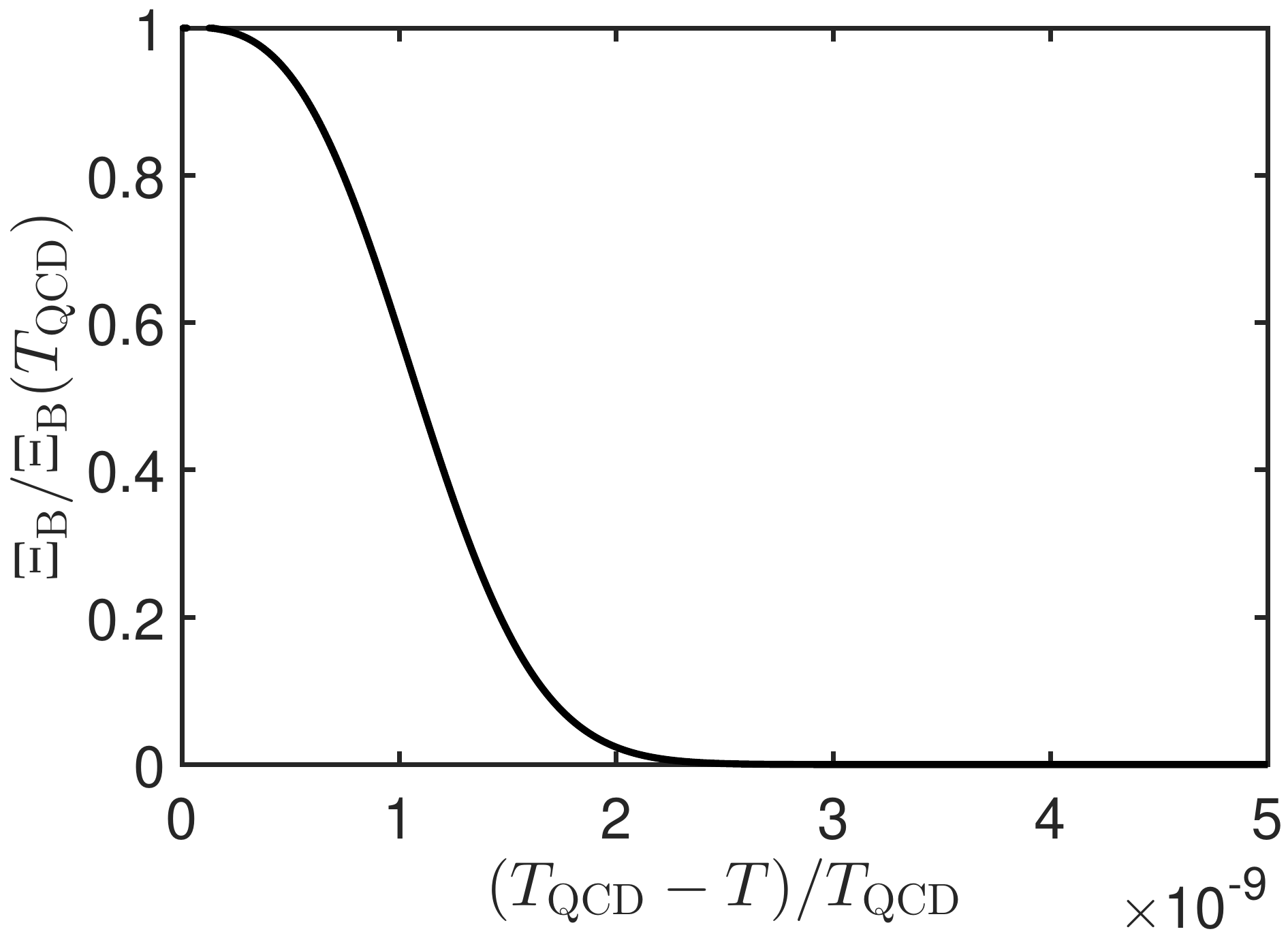}}
  \subfigure[]
  {\label{fig:Benergb}
  \includegraphics[scale=.35]{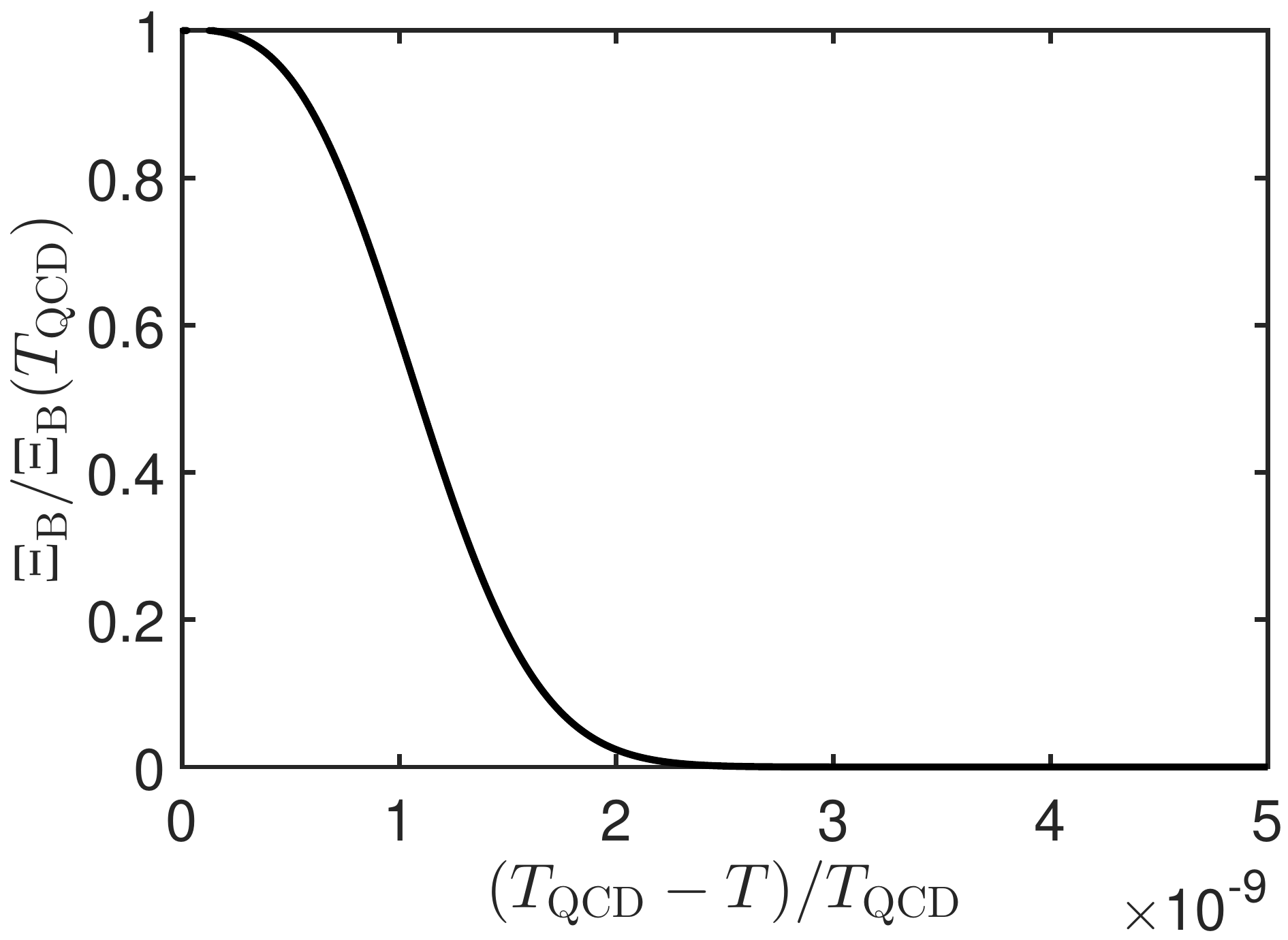}}
  \protect
\caption{The evolution of the total magnetic energy density $\Xi_{\mathrm{B}}\propto B_{+}^{(0)2}+B_{-}^{(0)2}$
normalized by its initial value. The parameters of the system are
the same as in Fig.~\ref{fig:Benerg}. Panel (a) corresponds to a
dense star; panel (b) to a dilute one.\label{fig:Benerg}}
\end{figure}

The magnetic fields in Fig.~\ref{fig:Baxions} decay quite rapidly.
It happens because of both the magnetic diffusion and the magnetic
field quenching in Eqs.~(\ref{eq:Bpeq}) and~(\ref{eq:Bmeq}). To
demonstrate this fact avoiding rapid oscillations visible in Fig.~\ref{fig:Baxions},
we show the evolution of the total magnetic energy $\Xi_{\mathrm{B}}\propto B_{+}^{(0)2}+B_{-}^{(0)2}$
in Fig.~\ref{fig:Benerg}. We normalize the total magnetic energy to its initial value $\Xi_{\mathrm{B}}(T_\mathrm{QCD}) \propto B_{+}^{(0)2}(T_\mathrm{QCD})$. Rapid oscillations are absent in Fig.~\ref{fig:Benerg}. The fact that, in the axion electrodynamics, the magnetic energy is a slowly varying function, unlike the magnetic field, was mentioned in Ref.~\cite{DvoSem20}. Again, both dense and dilute axion stars
are considered in Figs.~\ref{fig:Benerga} and~\ref{fig:Benergb}. We can see that these cases are almost identical.

\begin{figure}
  \centering
  \subfigure[]
  {\label{fig:axenerga}
  \includegraphics[scale=.35]{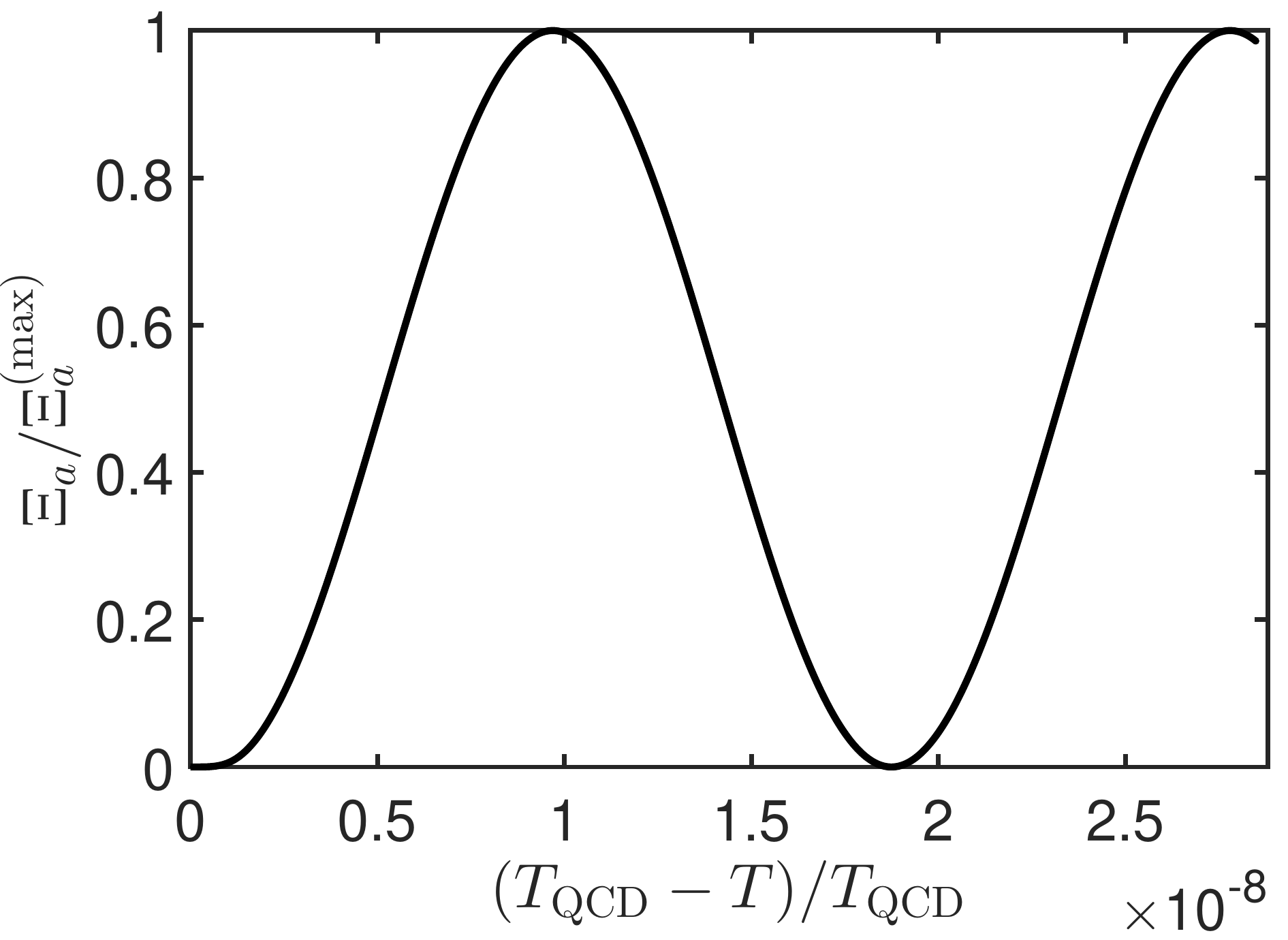}}
  \subfigure[]
  {\label{fig:axenergb}
  \includegraphics[scale=.35]{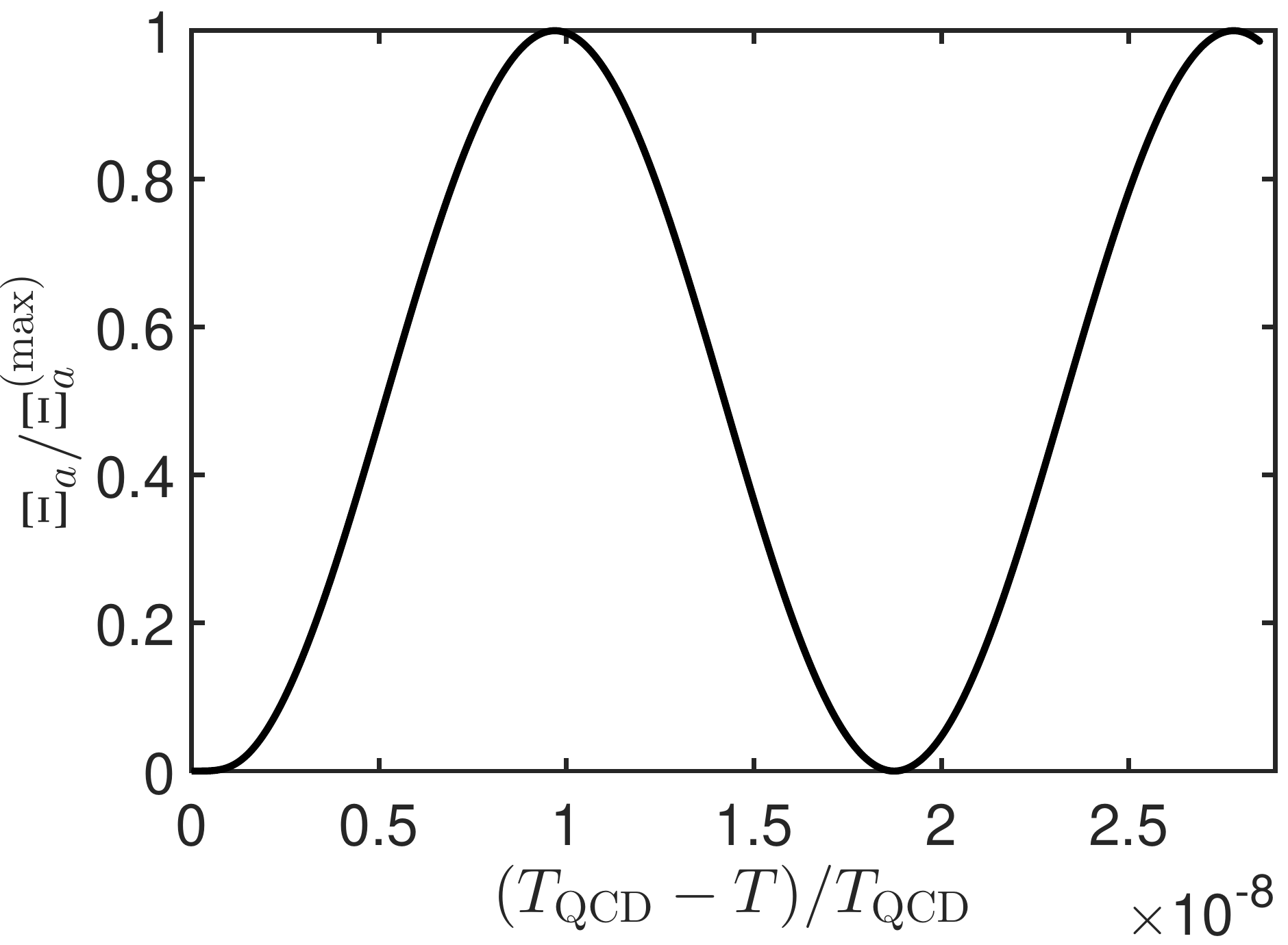}}
  \protect
\caption{The energy density $\Xi_{a}$ of $\varphi$, normalized by its maximal value,
vesus the plasma temperature. The parameters of the system are the
same as in Fig.~\ref{fig:Benerg}. Panel (a) corresponds to a dense
star; panel (b) to a dilute one.\label{fig:axenerg}}
\end{figure}

Finally, in Fig.~\ref{fig:axenerg}, we depict $\Xi_a$, which is the energy of $\varphi$
defined in Eq.~(\ref{eq:axen}). The value of $\Xi_a$ is shown for dense and dilute axion stars. We normalize $\Xi_a$ to its maximal value, which turns out to be great, $\Xi_{a}^{(\mathrm{max})}\gg 1$. The feature that $\Xi_{a}^{(\mathrm{max})}\gg 1$ results from the fact that rapidly oscillating magnetic fields in Fig.~\ref{fig:Baxions} transfer their energy to axions. It leads to the enhancement of the amplitude of axions energy oscillations even when magnetic fields decay. Such a behavior of the energy of $\varphi$ results from the term $\propto\left(\mathcal{B}_{+}^{2}+\mathcal{B}_{-}^{2}\right)/a^{2}$ in the right hand side of Eq.~(\ref{eq:dotPsi}). Note that the decoupling of the evolution of axions from magnetic fields was noticed earlier in Ref.~\cite{DvoSem20}. We also mention that $\Xi_a$ starts oscillating
with a frequency much lower than the magnetic field in Fig.~\ref{fig:Baxions}.


Unfortunately, it is technically difficult to trace the evolution
of the system at greater times numerically since we have two different
typical frequencies: one of them is related to magnetic field
oscillations (see Fig.~\ref{fig:Baxions}) another one to
oscillations of $\varphi$ (see Fig.~\ref{fig:axenerg}). The ratio of these frequencies is huge.

\section{Three dimensional model based on a spherical CS wave\label{sec:3D}}

In this section, we qualitatively consider the evolution of 3D magnetic fields under the influence of the inhomogeneous $\varphi$. Here, we also study the evolution of the system after the QCD phase transition when axions acquire a constant mass. We have demonstrated in Sec.~\ref{sec:1D} that the seed magnetic field decays quite rapidly; cf. Figs.~\ref{fig:Baxions} and~\ref{fig:Benerg}. That is why we neglect the conformal quantities and consider the physical magnetic fields.
	
By a spherical CS wave we means an analogue of CS waves in Eq.~\eqref{eq:Bpm}, which is defined in a compact domain rather than in the flat Euclidean space. The simplest example of a closed compact 3D domain is the standard 3D unite sphere, which is defined in 4D Euclidean space with the coordinates $(x_0,x_1,x_2,x_3)$ by the equation: 
\begin{equation}
  x_0^2 + x_1^2 + x_2^2 + x_3^2 = 1.	
\end{equation}
On the unite 3D sphere, we define right, $\B_+$, and left, $\B_-$, magnetic fields, as well as the following non-helical magnetic fields: $\B_\mathrm{A} = \B_+ - \B_-$ and $\B_\mathrm{B} = \B_+ + \B_-$. The magnetic field $\B_\mathrm{A}$ is not a complete analogue of the vector $\B_c$, given by Eq.~\eqref{eq:Bpm}, because the vector $\B_c$ is a right-polarized in the case $k>0$, whereas the vector $\B_\mathrm{A}$ is a mirror-symmetric. An interesting fact that $\B_\mathrm{A}$ is orthogonal to $\B_\mathrm{B}$ and the two magnetic modes are linked by the helicity integral. Let us remark that the magnetic vector $\B_+$ is well-known and was constructed in Ref.~\cite[Ex.~1.9, Ch.~III]{A-Kh}, as well as in Ref.~\cite{K}, using the Kelvin transformation described in Appendix~\ref{sec3}. The splitting
\begin{equation} 
  \B_+ = \frac{1}{2}[\B_\mathrm{A} + \B_\mathrm{B}],
\end{equation}
into the sum of two mirror-symmetric vectors is new.
	
Unfortunately, in a compact domain, we are not able the define a regular analogue of the gradient of the axion wavefunction $\b$ as in Eq.~\eqref{eq:Indeqgen}. The simplest analogue of the vector shift is the $k$-spectrum of the magnetic field. Instead of the evolution Eq.~\eqref{eq:Bpmeq} we get a nonlinear oscillator, determined by an infinite numbers of harmonics.  
	
We describe the analogue of Eq.~\eqref{eq:sysdim}. Then, we calculate the first-order derivative of a solution in the right-hand side of the equation
using the vector $\mathbf{b}$ in the left hand side of the equation. The axion density energy in the case is given analogously with Eq.~\eqref{eq:axconf}. This solution can be used to describe the evolution of the magnetic field in a short time, when an initial magnetic field is slowly varying, whereas the axion density energy changes rapidly. This situation can be implemented when we consider great effective axion mass $\mu = m \sigma_c/k^2$. If we change neither $m$ nor $k$, it can be achieved by considering great conductivity of plasma $\sigma_c$. Analogous situation was studied in Ref.~\cite{Dvo22}. Thus, here, we consider the situation opposite to that shown in Figs.~\ref{fig:Baxions} and~\ref{fig:axenerg}.
	
To define a spherical analogue of a CS wave we use standard MHD calculations on the Riemannian manifold given in Ref.~\cite[Def.~5.9, Ch.~1]{A-Kh}.
The standard 3D sphere of the unit radius is equipped by the following coordinate system $(\phi,\psi,\theta=2\chi)$, which is related with the Cartesian  coordinate system in  $\R^4$ by Eq.~(\ref{S}).
	
Let us define the curves $\Theta_\mathrm{A}$ and $\Theta_\mathrm{B}$ on $\mathbb{R}^4$ rather than on $\mathbb{C}^2$,
\begin{align}
  \Theta_\mathrm{A}(\phi; x_0, x_1, x_2, x_3)=\Theta_\mathrm{A}(\phi;x_0,x_1)  = &
  \left( R_\mathrm{A}\cos(\phi),  R_\mathrm{A}\sin(\phi),0,0 \right),
   \nonumber 
   \\
	\Theta_\mathrm{B}(\psi; x_0, x_1, x_2, x_3)  = \Theta_\mathrm{B}(\psi; x_2, x_3)  = & 
	\left(
		0,0,R_\mathrm{B}\cos(\psi), R_\mathrm{B}\sin(\psi) 
	\right).
\end{align}
Since $R^2_\mathrm{A}+R^2_\mathrm{B}=1$, let us put $R_\mathrm{A}=\sin(2\chi)$ and $R_\mathrm{B}=\cos(2\chi)$.
Thus, we can compute 
\begin{equation}\label{vect}
		\B_{\rm A} = \dd\Theta_\mathrm{A}/\dd\phi, \quad \B_{\rm B} = \dd\Theta_\mathrm{B}/\dd\psi.
\end{equation}
Using Eq.~\eqref{vect}, we define the associated differential one-forms on $\mathbb{R}^4$,
\begin{align}
  \label{f1}
		\beta_{\rm A}^{\rm R4} = &
		B^0_{\rm A} \dd x^0 + B^1_{\rm A} \dd x^1 = -\sin(\phi) R_\mathrm{A}\dd x^0 + 
		\cos(\phi) R_\mathrm{A} \dd x^1.
  \\
  \label{f2}
		\beta_{\rm B}^{\rm R4} = &
		B^2_{\rm B} \dd x^2 + B^3_{\rm B} \dd x^3 = -\sin(\psi)R_\mathrm{B}\dd x^2
		+ \cos(\psi)R_\mathrm{B} \dd x^3.
	\end{align}
We now define the mapping between points on the three-sphere $S^3$ and $\mathbb{R}^4$,
%
	\begin{eqnarray}\label{S}
		\Upsilon & = & (x_0, x_1, x_2, x_3), \nonumber \\
		x_0 & = & \cos(\phi)\cos(2\chi), \nonumber \\
		x_1 & = & \sin(\phi)\cos(2\chi), \nonumber \\
		x_2 & = & \cos(\psi)\sin(2\chi),  \nonumber \\
		x_3 & = & \sin(\psi)\sin(2\chi), 
	\end{eqnarray}
	with the coordinates of $S^3$: $\phi \in [0, 2\pi)$, $\psi \in [0, 2\pi]$,
	and $2\chi \in [0, \frac{\pi}{2}]$.
	We can now compute the differential one-forms $\beta_{\rm A}^{\rm R4}$ and $\beta_{\rm B}^{\rm R4}$ on $S^3$ as the pull-back
	under the mapping $\Upsilon$,
	\begin{align}
		\beta_{\rm A}^{\rm S3}  =  & \Upsilon^{*} \beta_{\rm A}^{\rm R4} 
		\nonumber
		\\
		& =
		\sin(\phi) \cos(2\chi) \sin(\phi) \cos(2\chi) \dd \phi + \sin(\phi) \cos(2\chi) \cos(\phi) \sin(2\chi) \dd 2\chi 	
		\nonumber
		\\
		& +
		\cos(2\chi)\cos(\phi)\cos(\phi)\cos(2\chi)\dd \phi -
		\cos(2\chi) \cos(\phi) \sin(\phi) \sin(2\chi) \dd 2\chi
		\nonumber 
		\\
		& =
		\cos^2(2\chi) \dd \phi,
		\label{4}
       \\      
		\beta_{\rm B}^{\rm S3}  = &
		\Upsilon^{*} \beta_{\rm B}^{\rm R4} 
		\nonumber
		\\
		& =
		\sin(\psi) \sin(2\chi) \sin(\psi) \sin(2\chi) \dd \phi - \sin(\psi) \sin(2\chi) \cos(\psi) \cos(2\chi) \dd 2\chi 
		\nonumber
		\\
		& +
		\cos(\psi)\sin(2\chi)\cos(\psi)\sin(2\chi)\dd \phi +
		\cos(\psi) \sin(2\chi) \sin(\psi) \cos(2\chi) \dd  2\chi 
		\nonumber
		\\
		& =
		\sin^2(2\chi) \dd \psi.
		\label{5}
\end{align}
The forms in Eqs.~\eqref{4} and~\eqref{5} obey the properties,
\begin{align}
  \label{6}
  \dd \beta_{\rm A}^{\rm S3}  =  & -2\cos(2\chi)\sin(2\chi) \dd  2\chi \wedge \dd \phi,
  \\
  \label{7}
  \dd \beta_{\rm B}^{\rm S3}  =  & 2\cos(2\chi)\sin(2\chi) \dd 2 \chi 
  \wedge \dd \psi. 
\end{align}
%
	We take their Hodge-dual $\star\dd\beta_{\rm A}^{S3}$,
		with the volume element $\dd V = \cos{2\chi}\sin(2\chi)  \dd\phi \wedge \dd 2\chi \wedge \dd\psi $,
	and find 
\begin{align}
  \label{D1}
		\star\dd\beta_{\rm A}^{S3}  = &
		\frac{1}{2}\sin^2(2\chi)\cos^2(2\chi)  \dd\psi = 2\sin^2(4\chi) \dd \psi,
  \\
  \label{D2}
		\star\dd\beta_{\rm B}^{S3}  = &
		-\frac{1}{2}\sin^2(2\chi)\cos^2(2\chi)  \dd\phi = 2\sin^2(4\chi) \dd \phi. 
	\end{align}
Finally, we get the result,
	\begin{align}
	\label{analog1}
	\dd\star\dd\beta_{\rm A}^{\rm S3} = & 2\dd \beta_{\rm  B}^{\rm S3},
	\\
   \label{analog2}
	\dd\star\dd\beta_{\rm B}^{\rm S3} = & 2 \dd \beta_{\rm  A}^{\rm S3}.
\end{align}
	

Let us define  smooth magnetic fields
\begin{align} 
  \B_+ = & \B_\mathrm{A} + \B_\mathrm{B} = \ast \dd \beta^{S3}_\mathrm{A} + \ast \dd \beta^{S3}_\mathrm{B},
  \notag
  \\
  \B_- = & \B_\mathrm{A} - \B_\mathrm{B} = \ast \dd \beta^{S3}_\mathrm{A} - \ast \dd \beta^{S3}_\mathrm{B},
\end{align}
on $S^3$, where
$\ast$ means a vector field, which is associated with the corresponding $2$-form by means of the volume form. 
The quantities
$\B_\mathrm{A}$ and $\B_\mathrm{B}$ are defined using the corresponding $2$-form in Eqs.~(\ref{6}) and~(\ref{7}).  Let us define a rescaling, $\B_\mathrm{A} \mapsto \frac{1}{2}\B_\mathrm{A}$ and $\B_\mathrm{B} \mapsto \frac{1}{2}\B_\mathrm{B}$, for simplicity. 

Using 
the coordinates $(\phi,\psi,2\chi=\theta)$ on $S^3$, we get that
\begin{equation}\label{vect2}
	\B_\mathrm{A}=(\sin\theta,0,0),
	\quad
	\B_\mathrm{B}=(0,\cos\theta,0).
\end{equation}
Obviously,
$\curl(\B_+)=2\B_+$ and $\curl(\B_-)=-2\B_-$.

\section{Harmonics\label{sec:HARM}}

Let us rescale the latitude
$\theta$ into $\chi$ by the formula,
\begin{equation}
  \theta = 2\chi, \quad \theta \in \left[0,\frac{\pi}{2}\right].
\end{equation}
Then, we define the magnetic harmonics
$\B_{1,A}$, which are called poloidal harmonics. A poloidal harmonic equals to zero at $\theta =0$.


It is convenient to remark for the calculations that the absolute value of the vectors
$\B_\mathrm{A}\cos(\theta)$ and $\B_\mathrm{B}\sin(\theta)$ coincides with functions of  $\theta$-coordinate and is defined as  $\cos(\theta)\sin(\theta)$. Directions of the vectors are perpendicular at an arbitrary point on $S^3$. 

Let us define the field distribution $\varphi(\theta,t)$, which is a pseudoscalar analogue of Eq.~\eqref{eq:axconf}, 
on the sphere. It also depends on time by the formula,
\begin{equation}\label{f}
	\varphi(\theta,t) = a_0\cos(n\theta)\sin(\omega t),
\end{equation}
where the parameter $a_0$ is assumed to be small.

Since $(\B_\mathrm{A} \cdot \B_\mathrm{B})=0$, the function $\varphi(\theta,t)$
obeys the equation, 
\begin{eqnarray}\label{eq7}
	\ddot{\varphi} - \nabla^2\varphi + m^2\varphi = 0,
\end{eqnarray}
and the following expression is valid: 
\begin{eqnarray}\label{eq8}
-\omega^2 + n^2 + m^2 = 0. 
\end{eqnarray}
We can normalize Eq.~(\ref{eq7}) by $\nabla^2$ and get that $\omega$ and $m$ in Eq.~(\ref{eq8}) are dimensionless. 
Note that Eq.~\eqref{eq7} is analogous to Eq.~\eqref{eq:axconf} where we neglect the universe expansion.

Let us consider the following sequence of transformations:
\begin{eqnarray}\label{trans}
	\begin{array}{c}
		\B_\mathrm{B} \stackrel{\curl}{\longrightarrow} 
		2\B_\mathrm{A} \stackrel{\times \nabla\cos(n\theta)}{\longrightarrow} -2n\sin(n\theta)\sin(\theta)\cos^{-1}(
		\theta)\B_\mathrm{B}\\ \stackrel{\curl}{\longrightarrow} 2n\frac{\mathrm{d}}{\mathrm{d}\theta}[\tan(\theta)\sin(n\theta)]\cot(\theta)\B_\mathrm{A} - 4n\tan(\theta)\sin(n\theta)\B_\mathrm{A}.
	\end{array} 
\end{eqnarray}
Based on Eq.~\eqref{trans}, we consider the case $n=2$. In this situation, one has
\begin{eqnarray}\label{trans2}
	\begin{array}{c}
		\B_\mathrm{B} \stackrel{\curl}{\longrightarrow} 2\B_\mathrm{A} \stackrel{\times \nabla\cos(2\theta)}{\longrightarrow}\\
		-8\sin^2(\theta)\B_\mathrm{B} \stackrel{\curl}{\longrightarrow} 16\cos^2(\theta)\B_\mathrm{A}  -16\sin^2(\theta)\B_\mathrm{A} =16\cos(2\theta)\B_\mathrm{A}.
	\end{array} 
\end{eqnarray}
Let us apply the calculations in Eq.~\eqref{trans2} to the formula,
\begin{eqnarray}\label{flow}
	\begin{array}{c}
		\dot{\B}_\mathrm{B} =
		a_0\sin(\omega t) \nabla \times
		[\nabla\cos(n\theta) \times (\nabla \times \B_\mathrm{B})] + \\
		\nabla \times[\alpha \B_\mathrm{B}] - \eta_m  \sin(\omega t) [\nabla \times [
		\nabla\times \B_\mathrm{B}]].
	\end{array}
\end{eqnarray}
where $\alpha$ is defined in Eq.~\eqref{eq:Indeqgen}.

Since $\alpha =  \omega\alpha_0 \cos(\omega t)\cos(n\theta)$,
for a small $\alpha_0$ only first-order terms are calculated. The first term is known from Eq.~(\ref{trans2}). The second term is given by $\nabla \times[\alpha \B_\mathrm{B}] = \omega \alpha_0  \cos(\omega t)  \nabla \times[\cos(n\theta)\B_\mathrm{B}]= \omega \alpha_0\cos(\omega t)[n\sin(n\theta)\sin^{-1}(\theta)\cos(\theta)\B_\mathrm{A} + 2\cos(n\theta)\B_\mathrm{A}] $.
For $n=2$ the second term is the following: $-2 \omega\alpha_0\cos(\omega t)[2\cos^2(\theta)-\sin^2(\theta)]\B_\mathrm{A}$.
The third term has the form, $-4\eta_m\sin(\omega t)\B_\mathrm{B}$.

To simplify the calculations we take $\eta_m=0$. In the case $n=2$, we have
\begin{equation}\label{flow2}
		\dot{\B}_\mathrm{B} =
		2a_0[8\sin(\omega t)\cos(2\theta) + \omega\cos(\omega t)(2\cos^2(\theta)-\sin^2(\theta))]\B_\mathrm{A}. 
\end{equation}
Let us transform Eq.~(\ref{flow2}) as
\begin{equation}\label{flow3}
	2a_0[8\sin(\omega t)  - \omega\cos(\omega t)]\cos(2\theta)\B_\mathrm{A} 
	+ 2\omega a_0\cos(\omega t) \cos^2(\theta) \B_\mathrm{A}.
\end{equation}
This calculation means that, in a stationary nonhelical magnetic field
$\B_\mathrm{B}$ with a fast oscillations of the pseudoscalar field, a first-order variation is given by a standing wave
represented by the second term in Eq.~(\ref{flow3}). The amplitude of the wave is determined by an amplitude of the pseudoscalar field. Additionally, we get a running wave in the first term in Eq.~(\ref{flow3}).

The parameter $a_0$ in Eq.~\eqref{flow3} is determined by the initial value of the axion field at the moment of the axion star formation according to Eq.~\eqref{f}. This parameter can be also related to the energy density of axion field as it is made in Sec.~\ref{sec:1D}; cf. Eq.~\eqref{eq:endensgen}. However, since the results of this section are general, i.e. they are applicable for both dense and dilute axion stars, we do not specify any particular value of $a_0$. The only assumption to be fulfilled in this section, is the small oscillations frequency of the magnetic field compared to that of the axion.

When the frequency
$\omega$ of $\varphi$ in Eq.~(\ref{flow2}) is great, the $\alpha$-effect,  which is related with the second term in Eq.~(\ref{flow2}) is dominated. Oppositely,
when a frequency $\omega$ is small, 
the first term in this equation is dominated. In each of the two limited cases, a standing wave is presented.
A running wave is presented when the first and the second term in Eq.~(\ref{flow2})
are of a same order.

Let us prove that, at the initial stage of a nonlinear process, the magnetic energy flow
	is defined by a dispersion of the magnetic helicity (see Ref.~\cite{A-V}) rather than by a magnetic helicity.  
	Based on  the condition, $\dot{\varphi}=0$, only harmonics $\sim \cos(\omega t)$ are presented. An elementary harmonic should be even, because of the boundary conditions, 
\begin{equation}
	\frac{\mathrm{d}}{\mathrm{d}\theta} \varphi(\theta,0)=0, 
\end{equation}	
for $\theta = 0$ and $\theta = \frac{\pi}{2}$.
	By the meaning of the problem, only spatial inhomogeneous axions are considered. Therefore the harmonic with $n=0$ in 
	Eq.~(\ref{f}) is absent.
	Based on this assumption, we get that
	\begin{equation}\label{d}
		\int \varphi(\theta,0) \mathrm{d} \theta = 0.
	\end{equation}
	
	Let us consider a contribution of the first term in Eq.~\eqref{flow}.
	One can prove that the magnetic helicity is conserved. Indeed,
\begin{multline}	
    \int (\nabla \times
	[\nabla\cos(n\theta) \times (\nabla \times \B_\mathrm{B})] \cdot \B_\mathrm{B}) \mathrm{d}S^3=
	\int ([\nabla\cos(n\theta) \times (\nabla \times \B_\mathrm{B})] \cdot \B_\mathrm{A}) \mathrm{d}S^3
	\\
	=\int_0^{\frac{\pi}{2}} \frac{\mathrm{d}}{\mathrm{d} \theta} [\cos(n \theta)] \mathrm{d} \theta = 0, \quad  n = 0 \pmod{2},
\end{multline}	
		where $\mathrm{d}S^3$ is the standard volume form on the unite sphere, which was considered above by means of  the coordinates $(\phi,\psi,\theta=2\chi)$.
		
	Let us calculate the density of the quadratic magnetic helicity flow because of the first term in Eq.~(\ref{trans2}). The second term, evidently, determines a linear growth of the magnetic helicity. To demonstrate this effect we put for simplicity $\alpha = 0$ and $\eta_m=0$. In Eq.~\eqref{trans}, let us consider $n=2$ by Eq.~(\ref{trans2}). 
At $t=0$, the first term in  Eq.~\eqref{trans} is given by
	\begin{equation}\label{flow2}
		\dot{\B}_\mathrm{B} =
		16a_0\cos(2\theta)\B_\mathrm{A}. 
\end{equation}

By this calculation we see that the magnetic energy flow exists with zero the magnetic helicity flow. One can see that there are two zones in the domain with different signs of the magnetic helicity flow. In this example, the magnetic energy relates with a spatial fluctuation of the magnetic helicity density (asymptotic ergodic Hopf invariant in Ref.~\cite{A-Kh}) rather than with the magnetic helicity. A local expression for the dispersion of the density variation is investigated in Ref.~\cite{A-V}.

\section{Discussion\label{sec:CONCL}}

In the present work, we have studied the simultaneous evolution of magnetic and spatially inhomogeneous pseudoscalar fields. The coordinate dependence of these fields have been chosen in a specific way imitating realistic situations taking place in astrophysical objects, e.g., in axion stars. We did not study the formation of the fields configuration. Our main goal was to analyze the time evolution of both magnetic and pseudoscalar fields.

We have started in Sec.~\ref{sec:AXELECTR} with writing down the main equations of the electrodynamics in the presence of a pseudoscalar field. We have obtained the main Eq.~\eqref{eq:Indeqgen} for the magnetic field evolution in the presence of a spatially inhomogeneous $\varphi$. Equation~\eqref{eq:Indeqgen} is a modification of the induction Eq.~\eqref{eq:Faraday} known in MHD. Our subsequent studies have been based on Eq.~\eqref{eq:Indeqgen}.

Two main cases have been considered. First, in Sec.~\ref{sec:1D}, we have studied the simplified 1D model where two CS waves with independent polarizations have been present. The pseudoscalar field was linearly decreasing within the crust of an axion star. The main impact of the nonzero gradient of $\varphi$ in this geometry is the mixing of independent CS waves. It does not contribute to the magnetic field instability.

The formation of axion and boson stars happens, e.g., owing to the gravitational interaction between particles~\cite{LevPanTka18}. We can approximate the distribution of pseudoscalar particles inside the star by Eq.~\eqref{eq:linearax}. When such a star is formed, we suppose that a magnetic field, with the strength in Eq.~\eqref{eq:Bpm}, appears inside. Then, we explore the evolution of $\varphi$ and the magnetic field on the basis of the initial conditions in Eqs.~\eqref{eq:Bpm} and~\eqref{eq:linearax}. We have found that the time scale of the magnetic field variation is quite short. Hence, we can neglect the gravitational interaction between axions, responsible for Eq.~\eqref{eq:linearax}, in Eq.~\eqref{eq:axconf}. Therefore, we do not require that, e.g., Eq.~\eqref{eq:linearax} obeys Eq.~\eqref{eq:axconf}. Hence, Eq.~\eqref{eq:linearax} is the given initial condition in our analysis. If we replace the simplified dependence of $\varphi$ on coordinates in Eq.~\eqref{eq:linearax} with more realistic smooth functions considered, e.g., in Ref.~\cite{Vic21}, it will just quantitatively modify the results.

In Sec.~\ref{sec:1D}, we have derived the system of ordinary differential equations for the amplitudes of independent CS waves. The system of differential Eqs.~\eqref{eq:Bpeq}-\eqref{eq:dotPsi} has been solved numerically. We have obtained that the magnetic field decays quite rapidly. Additionally, we have found that the evolution of dense and dilute axion stars is practically identical. Interacting axions and magnetic fields were supposed to exist in the early universe after the QCD phase transition.

Thus, if an axion star having magnetic fields with the proper configuration are created in the early universe just after the QCD phase transition, these fields will effectively disappear. Therefore, the magnetic traces of such axion stars are unlikely to reach present times. Nevertheless, we have obtained that the magnetic fields can transfer their energy to axions, enhancing their oscillations even when magnetic fields vanish. It means that one would have axion stars with rather great axion energy density. If such axion stars survive to later evolution times of the universe, they can reveal themselves in collisions with other astrophysical objects like neutron stars or black holes. Some observational consequences of such events are reviewed in Ref.~\cite{BraZha19}.

The change of the polarization of an electromagnetic wave in the presence of an axion field was studied in Ref.~\cite{Jai02}. In describing of that phenomenon, one accounts for the inhomogeneity of the axion wavefunction only in Eq.~\eqref{eq:Maxcov3} (see, e.g., Ref.~\cite{Raf96}). Whereas the term containing $\nabla \varphi$ in Eq.~\eqref{eq:Maxcov1} is neglected in Ref.~\cite{Raf96}. We take into account the inhomogeneity of $\varphi$ in both Eqs.~\eqref{eq:Maxcov1} and~\eqref{eq:Maxcov3}.

The nonzero $\nabla \varphi$ in Eq.~\eqref{eq:Maxcov1} was accounted for in the linear approximation in Ref.~\cite{HarSik92} to study cosmic birefringence in an axion gas. It leads to the rotation of the polarization plane of an electromagnetic wave. However, in our work, we examine how a nonzero $\nabla \varphi$ acts on the evolution of large scale magnetic fields rather than on electromagnetic waves. Therefore the mixing between two CS waves in the presence of inhomogeneous axions, considered in our work, is different from the phenomena discussed in Refs.~\cite{Jai02,HarSik92}.

Then, in Secs.~\ref{sec:3D} and~\ref{sec:HARM}, we have qualitatively studied more complicated situation based on the on the Hopf fibration. 
In this approach, Eq.~(\ref{eq:axconf}) in a first-order approximation of the solution is satisfied. However, the analogue of the evolution Eq.~(\ref{eq:sysdim})
is different and higher terms of a solution depend on lower ones.
Our main result is that, in the chosen geometry, the inhomogeneity of $\varphi$ leads to the change of the $\alpha$-dynamo parameter.
 
We have discussed inhomogeneous structures composed of pseudoscalar particles appearing after the QCD phase transition. Thus $\varphi$ is likely to correspond to a QCD axion which has the constant mass at this epoch. Moreover, we use the constraint on the QCD axion mass from Ref.~\cite{Bus22}. Nevertheless, our results can be applied to any axion like particles. One should just perform numerical simulations of Eqs.~\eqref{eq:Bpeq}-\eqref{eq:dotPsi} with different coefficients. The results of Secs.~\ref{sec:3D} and~\ref{sec:HARM} are general.

In summary, we have found that the impact of the nonzero gradient of the pseudoscalar field on the magnetic field evolution strongly depends on the geometry of the system. Of course, the comprehensive analysis should involve the simultaneous solution of both Eqs.~\eqref{eq:Indeqgen} and~\eqref{eq:axconf}. We plan to tackle this problem in one of our forthcoming works.

\section*{Acknowledgments}

We are thankful to E.~Maslov for useful discussions.

\appendix

\section{The Kelvin transformation\label{sec3}}

The Kelvin transformation is assumed to be a stereographic projection of the sphere $S^3$, outside a marked point $pt$, into the Euclidean space $\R^3$. This transformation is conformal, i.e. an angle between two vectors is unchanged. 
In Ref.~\cite{K}, the Kelvin transformation was
used to construct the MHD soliton in the Euclidean space with regular conditions at the infinity.

It is remarkable that the Kelvin transformation
$T: S^3 \setminus \{pt\} \to \R^3$
does not modify  equations. In the case the marked point $pt$ is the north pole on the sphere:
$\psi=\theta=0$, the magnetic mode $\B_\mathrm{A}$ is translated into a (generalized) toroidal mode $\BB_\mathrm{A}$, and the magnetic mode $\B_\mathrm{B}$ is translated into a (generalized) poloidal $\BB_\mathrm{B}$.   We will calculate only the transformation for the magnetic modes.

The Kelvin transformation $T$ is defined by the expression,
\begin{equation}\label{T}
	(\phi,\theta,\psi) \mapsto (r=\cot(\theta),\Psi,\phi),
\end{equation}
where, in the target, a spherical coordinate system with the latitude
$\Psi$ and the longitude $\phi$ is considered. The image of the $\psi$-coordinate  is calculated explicitly. It is independent of the $\phi$-coordinate in the target.
The absolute value of the gradient of the function
$\cot(\theta)$ determines the module of the conformal transformation $T$. This scale factor, which is a function in the target, is denoted by $K$.

The Kelvin transformation transforms a magnetic field
$\B$, which is a divergent-free field, on the source sphere into the corresponding magnetic field   $\BB= \frac{T_{\ast}(\B)}{K^3}$ on the target Euclidean space,  where
$K$ is the scale factor defined above. This function has an asymptotic $\propto r^{2}$, where $T_{\ast}$ is the translation of the vector by means of the differential of $T$. 
The vector-field $\BB$ is a magnetic field, because the flow of $\BB$ trought a surface $T(\Sigma) \subset \R^3$
equals to the flow of $\B^{S3}$ trought the surface $\Sigma \subset S^3$.

One can see that the mode
$\BB_\mathrm{B} = T_{\ast}(\B_\mathrm{B})K^{-3}$ is pointed
along the longitude $\phi$ and looks like a toroidal magnetic mode. The mode   $\BB_\mathrm{A}=T_{\ast}(\B_\mathrm{A})K^{-3}$ is pointed in a perpendicular direction along $(r,\Psi)$--coordinates and looks like a poloidal one. The scale factor
$K$ is related to the magnetic volume form  $K^3 \dd x$ in $\R^3$, which has an asymptotic  $\propto r^6$ along the radius. 
Both modes $\BB_\mathrm{A}$ and $\BB_\mathrm{B}$ have the asymptotic $\propto r^{-4}$.

The operator $\curl$ in $\R^3$ can be rewriten in the following form:
\begin{equation}
\curl(K^{-2}T_{\ast}(F)) = K^{-3}T_{\ast}(\curl_{S3}(F)),
\end{equation}
where the operator $\curl_{S3}$ and a vector-field $F$ are in the origine $S^3$. In particular, at the infinity, where
$K=K(r) \to +\infty$, the magnetic modes are almost potential and determine no currents. 

Magnetic helicity is kept by $T$. The helicity is calculated as an improper integral in the form,
\begin{equation}\label{chi}
  \chi_{\BB} =\int_{\R^3} (\AA \cdot \BB)\dd x = 
  \int_{\R^3} (K^{-2}T_{\ast}(\A),K^{-3}T_{\ast}(\B)) \dd x = \int_{S^3} ((\A \cdot \B)) \dd V = \chi_{\B},
\end{equation}	
where $\curl_{S3}(\A)=\B$. Here, we use the following formulas: $ K^3\dd V = \dd x $ and $ K^2((\dots,\dots)) = (\dots, \dots)$, which relate  the volume forms and  the scalar products in the source and in the target.

For $\B$ we have the well-defined boundary problem for the equation,
\begin{equation}\label{eq:curl}
\widehat{\curl} \widehat{\curl}  (\B) = \B, \quad \mathrm{div}(\B)=0, 
\end{equation}
where $\widehat{\curl} (\dots) = \curl(K^{-2}(\dots))$. Equation~\eqref{eq:curl} is the analog of the equations, $\Delta(\B) = 0$ and $\mathrm{div}(\B)=0$, in a infinite domain with $\B(r) \to 0$ and $r \to +\infty$
	for a non-uniformly prescribed magnetic permeability.
	In this case, we can consider for $\B$ a classifications problem  with quadratic boundary condition $| \B | = f(\theta,\varphi)$, which is a generalization of the classical Backus Problem\footnote{We are thankful to A.~Khokhlov for this remark.} (see Ref.~\cite{AP19} for more details).

\end{document}